\documentstyle[aps,twocolumn]{revtex}
\begin{document}
\draft
\title{Skyrme model and nonleptonic hyperon decays}
\author{G. Duplan\v{c}i\'{c}$^1$, H. Pa\v{s}agi\'{c}$^2$, M.
Prasza{\l}owicz$^3$, and
J.Trampeti\'{c}$^1$}
\address{$^1$Theoretical Physics Division, R. Bo\v{s}kovi\'{c} Institute, \\
P.O.Box 180, 10002 Zagreb, Croatia\\
$^2$Faculty of Transport and Traffic Engineering, University of Zagreb, \\
P.O. Box 195, 10000 Zagreb, Croatia\\
$^3$M. Smoluchowski Institute of Physics, Jagellonian University, \\
Reymonta 4, 30-059 Krak{\'o}w, Poland}
\date{\today}
\maketitle

\begin{abstract}
This report is an attempt to explain both s- and p-wave nonleptonic hyperon
decays by means of the QCD enhanced effective weak Hamiltonian supplemented
by the SU(3) Skyrme model used to estimate nonperturbative matrix
elements. The model has only one free parameter, namely, the Skyrme charge
$e$, which is fixed through the experimental values of the octet-decuplet
mass splitting $\Delta $ and the axial coupling constant $g_{A}$. Such a
dynamical approach produces nonleptonic hyperon decay amplitudes that agree
with experimental data reasonably well.
\end{abstract}


\narrowtext
In the Skyrme model, baryons emerge as soliton configurations of pseudoscalar
mesons \cite{sky}\nocite{adk,gua,yab}--\cite{wei}. Extension of the model to
the strange sector, in order to account for a large strange quark mass,
requires that appropriate chiral symmetry breaking terms are included.
The resulting effective Lagrangian can be treated by starting from a flavor
symmetric formulation in which the kaon fields arise from rigid
rotations of the classical pion field \cite{gua,NMPJW,MPSU3}. The associated
collective coordinates are canonically quantized to generate states that
possess quantum numbers of the physical strange baryons \cite{gua,wei,NMPJW}.
It turns out that the resulting collective Hamiltonian can be diagonalized
exactly even in the presence of the flavor symmetry breaking (SB) \cite
{yab}. This approach leads to a good description of hyperon masses, charge
radii, magnetic moments, etc. \cite{wei}. It should be noted that in the first
phenomenological applications of the Skyrme model one attempted to fit
absolute baryon masses, which required a ridiculously small pion decay
constant \cite{adk,MPSU3}. Nowadays it is understood that there exist 1/$N_{
{\rm c}}$ corrections to the total baryon masses that are not fully under
control and therefore only mass {\em splittings} can be reliably reproduced.
In this approach $f_{\pi }$ is kept at its experimental value. Hence the
results for nonleptonic hyperon decays
(NHD) \cite{tra} need to be updated accordingly.

Both s- and p-wave NHD amplitudes were quite successfully
predicted by using quark models with QCD enhancement factors \cite
{ga,do2,bur}. Note that there are not only current-algebra and ground-state
exchange pole-diagram terms, but there exist other important
contributions to both s and p waves. The so-called {\it factorizable}
contributions and/or kaon poles were estimated in \cite{ga,do2}.
Pole-\\
diagram contributions to p waves from the ($1/2^{+}$) -Rooper type of
resonances and to s-waves through the ($1/2^{-}$) -resonance
exchange were calculated in \cite{tra2}.

This report is an attempt to test whether the effective weak Hamiltonian
and the extended SU(3) Skyrme model are able to predict both s- and
p-wave NHD amplitudes. {\it The minimal number of couplings}
Skyrme model is used to estimate only the nonperturbative matrix elements of the
4-quark operators \cite{tra}. All remaining quantities entering the expressions
for the decay amplitudes like mass differences, coupling constants etc. are
taken from experiment. This approach uses only one free parameter, i.e.,
the Skyrme charge $e$. In order to avoid the unnecessary numerical burden,
throughout this report we use the {\it arctan ansatz} for the Skyrme profile
function \cite{dia}.

The starting point of our analysis of NHD
in the framework of the Standard Model \cite{ga} is the
effective weak Hamiltonian in the form of the current $\otimes $ current
interaction, enhanced by QCD. It is obtained by
integrating out heavy-quark and $W$-boson fields. This Hamiltonian contains
the 4-quark operators $O_{i}$ and the well-known Wilson coefficients\cite
{ga,do2}. For the most recent values, see Ref.\cite{bur}. For the
purpose of this paper, we use the Wilson coefficients from Ref.\cite{do2}: $
c_{1}=-1.90-0.61\zeta ,\;c_{2}=0.14+0.020\zeta
,\;c_{3}=c_{4}/5,\;c_{4}=0.49+0.005\zeta $, with $\zeta =V_{td}^{\ast
}V_{ts}/V_{ud}^{\ast }V_{us}$. Without QCD short-distance corrections, the
Wilson coefficients would be $c_{1}=-1,\;c_{2}=1/5,\;c_{3}=2/15$, and $
c_{4}=2/3$. In this paper we simply consider both possibilities and compare
the results.

Note that there exists a different approach of Ref.\cite{bije,sco2} in which
meson-baryon couplings are directly obtained from the chiral Lagrangian.
There, the effective phenomenological constants
extracted from experiment take into account all QCD effects hidden in the
structure of the effective Hamiltonian (including the {\it enhancement
factors} embodied in the values of the $c_{i}$ constants). This approach
gives comparable results for the $s$- waves but fails for
the $p$- waves \cite{sco2}.

The techniques used to describe NHD ($
1/2^{+}\rightarrow 1/2^{+}+0^{-}$ reactions) are known as a modified
current-algebra (CA) approach. The general form is
\begin{eqnarray}
&&<\!\pi (q)B^{\prime }(p^{\prime })|H^{eff}_{w}|B(p)\!>\!
= \bar{u}(p')[{\cal A}(q)+\gamma_5 {\cal B}(q)]u(p)
\nonumber \\
&&=\!\frac{-i}{2f_{\pi }}
<\!B^{\prime }(p^{\prime })|\hat{H}_{w}|B(p)\!>|_{q=0}+{\cal P}(q)+{\cal S}(q)\,.
\end{eqnarray}
Here the first term is the CA contribution, the second is
the modified pole term, and the third is a term that vanishes in the
soft-meson limit. The ${\cal P}(q)$ term contains the contribution from the
surface term, the soft-meson Born-term contraction, and the baryon pole
term, which are combined in a well-known way \cite{ga,do2}. It represents a
continuation of the CA result from the soft-meson limit. Further
continuation is contained in the factorizable term ${\cal S}(q)$, which is
proportional to the meson four-momenta.

The parity-violating amplitudes ${\cal A}$ receive contributions ${A^c}$
from CA commutator terms, factorizable terms ${\cal S}(q)$, and
pole terms from the $(1/2^-)$ - resonance exchange. The main contributions to
the ${\cal B}$ amplitudes come from the baryon pole terms ${\cal P}(q)$,
including both the ground state and the radially excited states.

The current-algebra $A^c$ and baryon-pole $B^{\cal P}$
amplitudes are well known from the literature. They contain weak matrix
elements defined as $a_{BB^{\prime}} = <B^{\prime}|H_w^{PC}|B>$, which have
the following general structure:
\begin{equation}
a_{BB^{\prime}} =\sqrt{2}G_F V^*_{ud}V_{us}
<B^{\prime}|c_{i}O_{i}^{PC}|B>.  \label{8}
\end{equation}

The factorizable term ${\cal S}(q)$ is calculated by inserting vacuum
states. It is therefore a factorized product of two current matrix elements,
where the first two-quark current is sandwiched between baryon states, while
the second two-quark current is responsible for pion emission.

The CA and the baryon-pole terms contain the 4-quark
operator matrix elements, which are nonperturbative quantities.
This is exactly the point at which the Skyrme model can be used.
Each of the operators $O_{i}$ from (\ref{8}) contains four types
of operators, namely,
$\bar{d}u\bar{u}s\,,\bar{d}s\bar{u}u\,,\bar{d}s\bar{d}d,\,
\bar{d}s\bar{s}s,\,$ and takes the form of the product of two
Noether SU(3) currents, which can be found in Refs.\cite{tra,sco}.
In our calculations we use four operators ${\hat{O}}_{i}$. The
first of them is
\begin{equation}
{\hat{O}}_{1}
=\frac{1}{4}\bar{q}{_{L}}\gamma _{\mu }({\lambda _{1}}-i{\lambda _{2}}
)q_{L}\,\bar{q}{_{L}}\gamma ^{\mu }({\lambda _{4}+}i{\lambda }_{5})q_{L},
\end{equation}
where the SU(3) properties are expressed explicitly in terms of
the Gell-Mann $\lambda $-matrices. The connection with the
effective Hamiltonian operators $O_{i}$ is obvious.

In order to estimate the matrix elements entering (\ref{8}), we take
the SU(3) extended Skyrme Lagrangian \cite{wei,sco}:
\begin{eqnarray}
{\cal L} &=&{\cal L}_{SK}^{(1)}+{\cal L}_{SK}^{(2)}+{\cal L}_{SB}+{\cal L}_{WZ},
\label{1}\\
{\cal L}_{SK}^{(1)} &=&\frac{f^2_{\pi}}{4}\int d^{4}x \,Tr\left(\partial _{\mu
}U\partial ^{\mu }U^{\dagger }\right) ,\;\;\,\mbox{etc.}  \nonumber
\end{eqnarray}
where ${\cal L}_{SK}^{(1,2)}$, ${\cal L}_{WZ}$, and ${\cal L}
_{SB}$ denote the $\sigma $-model, Skyrme, symmetry breaking (SB), and
Wess-Zumino (WZ) terms, respectively. For $U(x)\,\in \,SU(2)$, the SB and
WZ terms vanish. The $f_{\pi }=93$ MeV is the pion decay constant. Here the
space-time-dependent matrix field $U(\vec{r},t)\,\in \,SU(3)$ takes the form
\begin{equation}
U(\vec{r},t)=A(t){\cal U}(\vec{r})A^{\dagger }(t),  \label{3}
\end{equation}
where ${\cal U}(\vec{r})$ is the SU(3) matrix in which the Skyrme
SU(2) {\it ansatz} is embedded:
\begin{equation}
{\cal U}(\vec{r})=\left(
\begin{array}{cc}
exp(i{\vec{\tau}\cdot }\vec{n}F(r)) & 0 \\
0 & 1
\end{array}
\right) .  \label{4}
\end{equation}
The time-dependent collective coordinate matrix $A(t)\,\in \,SU(3)$ defines
the generalized velocities $A^{\dagger }(t)\dot{A}(t)={\frac{i}{2}}
\sum_{\alpha =1}^{8}\lambda _{\alpha }\dot{a}^{\alpha }$
and the profile function $F(r)$ is interpreted as a chiral angle
that parametrizes the soliton.

In this work we use the {\it arctan ansatz} for $F(r)$
\cite{dia}:
\begin{equation}
F(r)=2\arctan \left[ \left( r_{0}/r\right) ^{2}\right] .  \label{5}
\end{equation}
Here $r_{0}$ - the soliton size - is the variational parameter and the
second power of $r_{0}/r$ is determined by the long-distance behavior of the
massless equations of motion. After rescaling $x=ref_{\pi }$, one obtains
 $r_{0}/r=x_{0}/x$. The quantity $x_{0}$ has the
meaning of a ${\it dimensionless}$ size of a soliton and it is
determined by minimizing the classical mass $E_{cl}$.
All relevant integrals involving the profile function
turn into an integral representation of the Euler beta functions,
which can be evaluated ${\it analytically}$. The accuracy of
this method with respect to the numerical calculations is of the
order of a few percent. 
In the chiral limit of the SU(2) Skyrme model, we obtain
$x_0=\sqrt{15}/4$ and the {\it arctan ansatz} reproduces nucleon static
properties well \cite{MPSU3,art}.
Moreover we gain an insight on how different quantities depend on
the soliton size, which in turn is a function of the symmetry
breaker and $e$.

In the SU(3) extended Lagrangian (\ref{1}) we have a new set of
parameters, namely, ${\hat{x}}=36.4$, ${\beta }^{\prime
}=-2.98\times 10^{-5}$GeV$^{2}$, ${\delta }^{\prime }=4.16\times
10^{-5}$GeV$ ^{4}$, determined from the masses and decay constants
of the pseudoscalar mesons \cite{wei}. Owing to the presence of
the $\beta ^{\prime }$ and $\delta ^{\prime }$ terms in $E_{cl}$,
$x_{0}$ becomes a function of $e$, $f_{\pi }$, $\beta ^{\prime }$,
and $\delta ^{\prime }$, and it is equal
\begin{equation}
x^{\prime }{}_{0}^{2}=\frac{15}{8}\left[ 1+\frac{6\beta ^{\prime }}{f_{\pi
}^{2}}+\sqrt{\left( 1+\frac{6\beta ^{\prime }}{f_{\pi }^{2}}\right) ^{2}+
\frac{30\delta ^{\prime }}{e^{2}f_{\pi }^{4}}}\;\right] ^{-1},
\end{equation}
where we use the symbol $x^{\prime }_{0}$ to distinguish it from
the SU(2) case. After introducing the SB terms into the Lagrangian
(\ref{1}), one can either treat them as a perturbation
\cite{MPSU3} or one can try to sum up the perturbation series by
numerically diagonalizing the resulting Hamiltonian \cite{yab}.

The fitting procedure employed in this work is based
on taking the physical values for $f_{\pi }$ and $f_{K}$ which
takes care of the  SU(3) symmetry breaking. In fact the SB
affects the calculations in three different ways: (a) through the
soliton size $x^{\prime}_0$; (b) via the explicit SB in the currents;
(c) through the admixture of the higher SU(3) representations 
in the baryon wave functions. It was shown in Ref.\cite{sco} that
the latter contributions to NHD are small, since the higher representations
enter with small weights. Our estimate shows that they are of the
order of 15 \%. This uncertainty is of the order of the accuracy of
the model which is reflected in the variation of the Skyrme parameter $e$
depending on which static property is used in the fitting procedure. 
In the remainder of this paper we use the SU(3)
symmetric wave functions.

For the evaluation of NHD, the important baryon static properties are the
octet-decuplet mass splitting $\Delta $ and the axial decay coupling
constant $g_{A}$. The value of the only free parameter $e\approx 4$
was successfully adjusted to the mass difference $\Delta $ of the
low-lying $1/2^{+}$ and $3/2^{+}$ baryons \cite{wei}.
However, if we fix $\Delta $, the constant $g_{A}$
is underestimated. This is a well-known problem of the Skyrme model,
which can be cured in the more sophisticated chiral models involving quarks
\cite{WaWa}.

Therefore, we determine two values of the charge $e$
through fixing $\Delta $ and $g_{A}$ to their experimental values.
The {\it arctan ansatz} gives
\begin{eqnarray}
\Delta &=&\frac{3}{2\lambda _{c}(x_{0}^{\prime })},  \label{delta} \\
g_{A} &=&\frac{14\pi }{15e^{2}}\left( 2x{^{\prime }}_{0}^{2}+\pi \right)
\nonumber \\
&&{}+(1-\hat{x})\frac{16\pi \beta ^{\prime }}{225e^{2}f_{\pi }^{2}}x{
^{\prime }}_{0}^{2}+\frac{7\sqrt{2}N_{c}}{192ef_{\pi }}\frac{x_{0}^{\prime }
}{\lambda _{s}(x_{0}^{\prime })}, \label{11}
\end{eqnarray}
where
\begin{eqnarray}
\lambda _{c}(x_{0}^{\prime }) &=&\frac{\sqrt{2}\pi ^{2}}{3e^{3}f_{\pi }}
\left[ 6 \left(1+2\frac{\beta ^{\prime }}{f_{\pi }^{2}}\right)x{^{\prime }}_{0}^{3}+
\frac{25}{4}x_{0}^{\prime }\right], \label{lc} \\
\lambda _{s}(x_{0}^{\prime }) &=&\frac{\sqrt{2}\pi ^{2}}{4e^{3}f_{\pi }}
\left[ 4 \left(1-2(1+2{\hat x})
\frac{\beta ^{\prime }}{f_{\pi }^{2}}\right)x{^{\prime }}_{0}^{3}+
\frac{9}{4}x_{0}^{\prime }\right]. \label{ls}
\end{eqnarray}
The quantity $\lambda _{c}(x_{0}^{\prime })$ represents the rotation
moment of inertia in coordinate space, while the $\lambda _{s}(x_{0}^{\prime})$
is the moment of inertia for flavor rotations in the direction of the
strange degrees of freedom, except for the eighth direction\cite{wei,MPSU3}.
The static kaon fluctuations were omitted\cite{flu}
in the derivations of equations (\ref{delta} - \ref{ls}).

For the Lagrangian ${\cal L}$, we calculate the matrix element of the
product of two $(V-A)$ currents between the octet states
using of the Clebsch-Gordon decomposition \cite{tra}:
\begin{equation}
<B_{2}|{\hat{O}}^{(SK)}|B_{1}>=
\Phi^{SK} \times \sum_{R} C_R
\end{equation}
where $\Phi^{SK}$ is a dynamical constant and $C_R$ denote the
pertinent sum of the SU(3) Clebsch-Gordon coefficients in the
intermediate representation $R$. The total
matrix element is simply a sum $<{\hat{O}}_{i}^{(SK)}+{\hat{O}}
_{i}^{(WZ)}+{\hat{O}}_{i}^{(SB)}>$, with $i=1,\ldots ,4$. The
quantities $ \Phi $ are given by the overlap integrals of the
profile function. Using the {\it arctan anzatz} (\ref{5}), we
obtain analytical expressions for the integrals as functions of
$x_{0}^{\prime }$:
\begin{eqnarray}
{\Phi }^{SK} &=&3\sqrt{2}{\pi ^{2}}\left( 2x_{0}^{\prime }+\frac{15}{
2x_{0}^{\prime }}+\frac{847}{64}\frac{1}{x_{0}^{\prime 3}}\right) \frac{
f_{\pi }^{3}}{e},  \nonumber \\
\Phi ^{WZ} &=&\frac{231}{512}\frac{\sqrt{2}}{\pi ^{2}}\frac{1}{x_{0}^{\prime
3}}(ef_{\pi })^{3},  \nonumber \\
\Phi ^{SB} &=&\left( 1-{\hat{x}}\right) \beta ^{\prime }\frac{4{\pi }^{2}}{
\sqrt{2}}\left( x_{0}^{\prime }+\frac{45}{8x_{0}^{\prime }}\right) \frac{
f_{\pi }}{e}.  \label{10}
\end{eqnarray}
For the ${\hat{O}}_{1}$ operator, $R=8_{a,s}$ or $27$. ; then
\begin{eqnarray}
& &<p\uparrow |{\hat{O}}_{1}|\Sigma ^{+}\uparrow > = -\frac{1}{4}\left( {\frac{
2}{25}}|_{8}+{\frac{1}{675}}|_{27}\right) \Phi
^{SK} \label{11A} \\
& &{}-\frac{1}{4}\left( {\frac{2}{25}}|_{8}-{
\frac{1}{75}}|_{27}\right) \Phi ^{WZ}
-\frac{1}{4}\left( {\frac{7}{75}}|_{8}+{\frac{17}{1050}}|_{27}\right)
\Phi ^{SB}.  \nonumber
\end{eqnarray}
The $27$-piece is very small, which is an important proof of the octet
dominance.

By fixing $\Delta $ and $g_{A}$ to their
experimental values, we obtain $e=4.228$ and $e=3.385$, respectively. In
further calculations of the NHD amplitudes, we use the mean value $e=3.81$
\cite{del} and $x^{\prime}_0 |_{e=3.81}=0.8782$, i.e., 10\% less than in 
the massless case.
For $f_{\pi }=93$ MeV and $e=3.81$, we obtain the following
numerical values of the integrals (\ref{10}) in units of GeV$^{3}$:
\begin{equation}
\Phi ^{SK}=0.264,\;\;\Phi ^{WZ}=0.004,\;\;\Phi ^{SB}=0.005.  \label{phi}
\end{equation}
From eqs.(\ref{11A}) and (\ref{phi}) we find the following structure
for a typical matrix element:
\begin{eqnarray}
&<&p\uparrow |{\hat{O}}_{1}|\Sigma ^{+}\uparrow >=  \nonumber \\
&&(-20.37\Phi ^{SK}-16.67\Phi ^{WZ}-27.38\Phi ^{SB})10^{-3}=  \nonumber \\
&&(-5.38|_{SK}-0.06|_{WZ}-0.14|_{SB})10^{-3}GeV^{3}.  \label{12}
\end{eqnarray}
It is clear that on top of the octet dominance we also find the
dominance of the Skyrme Lagrangian currents over the WZ and
SB currents in the evaluation of a typical weak matrix
element between two hyperon states. For $e\approx 4$,
the SB and WZ terms are of comparable size and their coherent contribution
to (\ref{12}) is below $4\%$.
 We see therefore from (\ref{10}) that within this accuracy
the result for (\ref{12}) scales like $1/e$ (up to 5\% due to the 
small variations of the soliton size which weakly depends on $e$).
The change of $e$ between $3.385$ and $4.228$ produces $ 14\%$
variations  of the amplitudes $A^c(0)$ and $B^{\cal P}(m^2_{\pi})$
around their mean values given in  Table \ref{t:tab1}. 

In this work we have added factorizable, $A^{\cal S}(m^2_{\pi})$
and $B^{\cal S}(m^2_{\pi})$,
contributions to the Skyrme model amplitudes $A^c(0)$ and $B^{\cal P}(m^2_{\pi})$.
The complete results are given in Table \ref{t:tab1}. Comparison of the total
amplitudes ${\cal A}(m^2_{\pi})$ and ${\cal B}(m^2_{\pi})$
with experiment shows the following:

(a) Short-distance corrections to the effective weak Hamiltonian are beyond
doubt very important.

(b) Signs and order of magnitudes of all amplitudes are always correctly
reproduced.

(c) s waves are in good agreement with experiment.

(d) The Pati-Woo theorem violation\cite{pati} and the 27-contaminations are
found to be small. It is clear that the nonvanishing ${\cal A}(\Sigma^+_+)$
amplitude is still too small, in good accord with small values of the
27-contamination\cite{foot}, and that additional contributions are needed\cite
{tra2}.

(e) p waves are subject to some uncertainties. Namely, in the Ref.\cite
{bije} was shown that, in the Skyrme model, a contact term appeared and should
be added to the results for p-waves. That has been taken care of in Ref.
\cite{sco2} and is not present in our approach. In our opinion,
${\cal B}(m^2_{\pi})$ amplitudes are not fully described by our formulas; nevertheless,
they agree with experiment reasonably well.

(f) Finally, the factorizable contributions are small,
and represent the fine tuning to the total amplitudes.

To conclude, we would like to emphasize the fact that
the pure Skyrme model Lagrangian ${\cal L}$ cannot explain nonleptonic
hyperon decays\cite{sco2}. However, the QCD-corrected weak Hamiltonian $
H_{w}^{eff}$, together with the inclusion of other possible types of
contribution to the total amplitudes ($K$, $K^{\ast }$-poles, and/or
factorization; $(1/2^{{\pm}\ast })$-poles, etc.)
supplemented by the Skyrme model, leads to a correct answer.
This includes the explanation of the octet
dominance, the $|{{\Delta I}}|=1/2$ selection rule,
${\cal A}(\Sigma _{+}^{+})\not=0$, and the p/s-wave puzzle.
Nevertheless, this is certainly a matter of another series of studies.
\newline

One of us (J.T.) would like to thank J. Wess for many useful discussions and 
the Ludwig Maximillians Universit$\ddot a$t, M$\ddot u$nchen,
Section Physik, were part of
this work was done, for its hospitality.
This work was supported by the Ministry of Science and Technology of the
Republic of Croatia under the contract No. 00980102. M.P. was supported by
the Polish KBN Grant PB~2~P03B~{\-}019~17.

\renewcommand{\arraystretch}{1.4}
\begin{table}[tbp]
\caption{The s-wave (${\cal A}$) and p-wave (${\cal B}$) NHD
amplitudes. Choices (off, on) correspond to
the amplitudes without and with inclusion of short-distance corrections,
respectively. For the sake of
comparison, we have added the constituent quark-model evaluation of the $A^c$
and $B^{\cal P}$ amplitudes [9,10].}
\label{t:tab1}
\begin{tabular}{c|ccccccccccc}
$Amplitude (10^{-7}) $ &  &  &  & $(\Lambda^0_-) $ &  & $(\Xi^-_-) $ &  & $
(\Sigma^+_0) $ &  & $(\Sigma^+_+) $ &  \\ \hline\hline
$A^c (0) $ &  & $off $ &  & $2.02 $ &  & $-2.94 $ &  & $-2.28 $ &  & $0.02 $
&  \\
&  & $on $ &  & $3.84 $ &  & $-5.56 $ &  & $-4.34 $ &  & $0.04 $ &  \\ \hline
$A^{{\cal S}}(m^2_{\pi}) $ &  & $off $ &  & $0.03 $ &  & $-0.57 $ &  & $
-0.49 $ &  & $0 $ &  \\
$\cite{ga}$ &  & $on $ &  & $-0.42 $ &  & $0.25 $ &  & $-0.01 $ &  & $0 $ &
\\ \hline
${\cal A}(m^2_{\pi}) $ &  & $off $ &  & $2.05 $ &  & $-3.51 $ &  & $-2.77 $
&  & $0.02 $ &  \\
$(this \; work)$ &  & $on $ &  & $3.42 $ &  & $-5.31 $ &  & $-4.35 $ &  & $
0.04 $ &  \\ \hline
$Exp. \cite{rpp} $ &  &  &  & $3.35 $ &  & $-4.85 $ &  & $-3.27 $ &  & $0.13 $
&  \\ \hline
$A^c (0) $ &  & $off $ &  & $0.78 $ &  & $-1.86 $ &  & $-1.36 $ &  & $0 $ &
\\
$CQM \; \cite{ga}$ &  & $on $ &  & $1.49 $ &  & $-3.53 $ &  & $-2.59 $ &  & $
0 $ &  \\ \hline\hline
$B_{(1/2^+)}^{{\cal P}}(m^2_{\pi})$ &  & $off $ &  & $20.1 $ &  & $21.8 $
&  & $13.7 $ &  & $14.8 $ &  \\
&  & $on $ &  & $38.1 $ &  & $41.4 $ &  & $25.9 $ &  & $28.2 $ &  \\
\hline
$B^{{\cal S}}(m^2_{\pi}) $ &  & $off $ &  & $3.6 $ &  & $-1.5 $ &  & $-0.4 $
&  & $0 $ &  \\
$\cite{ga}$ &  & $on $ &  & $6.0 $ &  & $-2.4 $ &  & $0.4 $ &  & $0 $ &  \\
\hline
${\cal B}(m^2_{\pi}) $ &  & $off $ &  & $23.7 $ &  & $20.3 $ &  & $13.3 $
&  & $14.8$ &  \\
$(this \; work)$ &  & $on $ &  & $43.4 $ &  & $38.2 $ &  & $26.3 $ &  & $
28.2 $ &  \\ \hline
$Exp. \cite{rpp} $ &  &  &  & $22.3 $ &  & $17.4 $ &  & $26.6 $ &  & $42.2 $
&  \\ \hline
$B_{(1/2^+)}^{{\cal P}}(m^2_{\pi}) $ &  & $off $ &  & $2.9 $ &  & $7.8 $ &
& $7.3 $ &  & $10.4$ &  \\
$CQM \; \cite{ga}$ &  & $on $ &  & $5.6 $ &  & $14.8 $ &  & $13.9 $ &  & $
19.7 $ &
\end{tabular}
\end{table}
\renewcommand{\arraystretch}{1}

\end{document}